\documentstyle[twocolumn,epsf,psfig]{mn}
\title{Two-integral Schwarzschild models}
\author[E.K.~Verolme \& P.T~de Zeeuw]
{E.K.~Verolme\thanks{verolme@strw.leidenuniv.nl} and P.T.~de Zeeuw\\
Leiden Observatory, Postbus 9513, 2300 RA Leiden, The Netherlands\\
}
\date{Accepted 0000 0000.
      Received 0000 0000;
      in original form 0000 0000}
\pagerange{\pageref{firstpage}--\pageref{lastpage}}
\pubyear{0000}
\begin{document}
\maketitle
\label{firstpage}
\begin{abstract}
We describe a practical method for constructing axisymmetric
two-integral galaxy models (with distribution functions of the form
$f(E,L_z)$, in which $E$ is the orbital energy, and $L_z$ is the
vertical component of the angular momentum), based on Schwarzschild's
orbit superposition method. Other $f(E,L_z)$-methods are mostly based
on solving the Jeans equations or on finding the distribution function
directly from the density, which often places restrictions on the
shape of the galaxy. Here, no assumptions are made and any axisymmetric
density distribution is possible.  The observables are calculated
(semi-)analytically, so that our method is faster than most previous,
fully numerical implementations. Various aspects are tested
extensively, the results of which apply directly to three-integral
Schwarzschild methods. We show that a given distribution function can
be reproduced with high accuracy and investigate the behaviour of the
parameter that is used to measure the goodness-of-fit. Furthermore, we
show that the method correctly identifies the range of cusp clopes for
which axisymmetric two-integral models with a central black hole do
not exist.
\end{abstract}

\begin{keywords} galaxies: elliptical and lenticular, cD - galaxies:
kinematics and dynamics - galaxies: structure
\end{keywords}

\section{Introduction}
\label{sec:intro}

The fundamental quantity in galaxy dynamics is the distribution
function (DF), which specifies the distribution of the stars in a
galaxy over position and velocity. By Jeans' theorem, the DF is a
function of the isolating integrals that are conserved by the
corresponding potential (e.g., Lynden-Bell 1962; Binney 1982).
Axisymmetric galaxies, which we will study here, conserve at least the
two classical integrals of motion, the energy $E$ and the vertical
component of the angular momentum $L_z$. The DF can generally not be
measured directly, but since observationally accessible quantities,
such as the projected density and the line-of-sight velocity, are
simple moments of the DF, photometric and kinematic observations can
provide information on its properties. In some cases, the part of
a two-integral distribution function that is even in the velocities, 
$f(E,L^2_z)$, can be obtained analytically by using integral transforms 
to solve the relation between the DF and the intrinsic density $\rho(R,z)$, 
where $(R,\phi,z)$ are the usual cylindrical coordinates. The odd part can
be found similarly from $\rho \langle v_\phi \rangle$, where $\langle
v_\phi \rangle$ is the mean streaming velocity.  These Laplace
(Lynden-Bell 1962; Lake 1981), Stieltjes (Hunter 1975b) and
Laplace-Mellin (Dejonghe 1986) transforms have the drawback that
numerical implementation is difficult and that they require $\rho$ and
$v_\phi$ to be explicit functions of the potential $V$ and the
cylindrical radius $R$. A more general formalism (Hunter \& Qian 1993;
Qian et al.\ 1995, hereafter Q95), the HQ-method, uses contour
integration instead of integral transforms. This means it is simpler
to implement, does not explicitly require $\rho(R,V)$ and
$v_\phi(R,V)$, but a suitable contour for the integration has to be
chosen, which may not always be at hand.

Because of the drawbacks of these analytical formalisms, numerical
methods that can be applied to arbitrary potential-density pairs are
more attractive. Various methods have been developed, many of which
circumvent knowledge of the DF by solving the Jeans equations directly
(van der Marel et al.\ 1994; Magorrian 1995), while others assume that
the DF can be represented by a superposition of basis functions
(Dehnen \& Gerhard 1994; Kuijken 1995). Accurate estimates of the
mass-to-light ratio $M/L$ can be obtained in both ways (van der Marel
1991; Shaw et al.\ 1993; van den Bosch 1996). The predicted central
black hole masses that are obtained with two-integral models
(Magorrian et al.\ 1998) seem to overestimate the true values, but are
still very useful to narrow down the parameter range for more
sophisticated modeling (van der Marel et al.\ 1998, hereafter vdM98;
Gebhardt et al.\ 2000; Bower et al.\ 2001).

Numerical orbit integrations and observations of, e.g., the anisotropy
of the stellar dispersions in the solar neighbourhood show that, in
addition to $E$ and $L_z$, a third integral is conserved for most
orbits. In separable potentials, this third integral is exact and has
a closed form (e.g., Kuzmin 1956; de Zeeuw 1985), allowing a direct
calculation of the DF (e.g., Dejonghe \& de Zeeuw 1988).  The most
general family of these potentials corresponds to flattened mass
models with constant-density cores (de Zeeuw, Franx \& Peletier 1986),
which provide a poor description of the inner regions of most
galaxies, since these contain a central density cusp (Lauer et al.\
1995). We conclude that many of the existing two-integral, as well as
the analytical three-integral methods, are applicable to a limited
range of galaxy models.

A flexible method for calculating numerical galaxy models was designed
by Schwarzschild (1979, 1982), who represented the DF numerically by
the occupation numbers in a superposition of building blocks. There
are no restrictions on the density or the potential, and no a priori
assumptions have to be made about the shape or the degree of
anisotropy of the galaxy. Schwarzschild's original implementation was
aimed at reproducing a given triaxial density distribution and was
subsequently applied to a large variety of galaxy models, from
spherically and axially symmetric (Richstone \& Tremaine 1984; Levison
\& Richstone 1985) to triaxial density distributions (e.g., Vietri
1986; Statler 1987; Schwarzschild 1993, Merritt \& Fridman 1996;
Siopis \& Kandrup 2000).  More general versions of the method can also
reproduce kinematical observations of spherically and axially
symmetric galaxies that obey up to three integrals of motion (Zhao
1996; Rix et al.\ 1997, hereafter R97; Cretton et al.\ 1999, hereafter
C99; Cretton et al.\ 2000; H\"afner et al.\ 2000).\looseness=-2

In most implementations of Schwarzschild's method, orbits are used as
individual building blocks. In non-separable potentials the third
integral of motion is not known explicitly, so that the orbital
properties can only be obtained by solving the equations of motion
numerically. This orbit integration has to be carried out until the
relevant quantities have averaged out and do not change upon a new
time-step (Pfenniger 1984; Copin et al.\ 2000). Especially for orbits
that are near-stochastic, or that have nearly commensurate fundamental
frequencies, this condition is difficult to achieve and the orbit
integration can be very time-consuming. Furthermore, an orbit fills a
region in space that can have sharp edges, depending on the
combination of integrals of motion that it obeys. A superposition of a
finite (and relatively small) number of orbits can therefore show
artefacts that are caused by these edges. One way to solve this is to
`blur' orbits in phase-space randomly, another is to use building
blocks that are smoother (e.g., Zhao 1996; Merritt \& Fridman 1996;
R97; C99). Of particular interest are the so-called isotropic and
two-integral components (ICs and TICs). Isotropic components, 
which are completely specified by their energy $E$, fill the volume
inside the equipotential at $E$, while TICs, which are fixed once 
$E$ and $L_z$ are chosen, completely fill the corresponding 
zero-velocity curve (ZVC). The advantages of these building blocks is
apparent: both the equipotential and the ZVC are smooth surfaces, and 
ICs and TICs can be considered as weighted combinations of all orbits with 
the same energy $E$ (or $E$ and $L_z$), including the irregular orbits. 

These building blocks can be used in Schwarz\-schild's method in 
two ways. They can be added to every energy or every combination 
$(E,L_z)$ of a (three-integral) orbit library, which will `automatically' 
take care of the stochastic orbits (Zhao 1996; R97; C99). Alternatively, 
a fully isotropic or two-integral model can be constructed by using 
only ICs or TICs as building blocks. C99 suggested that this might be a 
practical way of constructing such models, and derived some analytic 
properties, but did not pursue these models further. 
We do so in this paper and concentrate on the two-integral case, since 
the properties of the isotropic components follow from those of the TICs 
by taking $L_z=0$.

In \S\ref{sec:method}, we collect the properties of the TICS, and
summarize how we use them in Schwarzschild's method. The numerical
aspects are discussed in detail in \S\ref{sec:implementation}.  We
show in \S\ref{sec:examples} that our method can reproduce a model
with a known analytical distribution function with high accuracy.  We
also study a set of mass models with a central black hole, and show
that our method is able to detect when a self-consistent solution
does not exist. We summarize our conclusions in \S\ref{sec:summary}.
Applications of our software to model the observed two-dimensional
surface brightness and kinematics of nearby elliptical galaxies
observed with {\tt SAURON} and {\tt STIS} are described in two
follow-up papers (Verolme et al.\ 2002, in preparation; 
McDermid et al.\ 2002, in preparation).

\section{Two-integral components}
\label{sec:method}

\begin{figure}
{\psfig{file=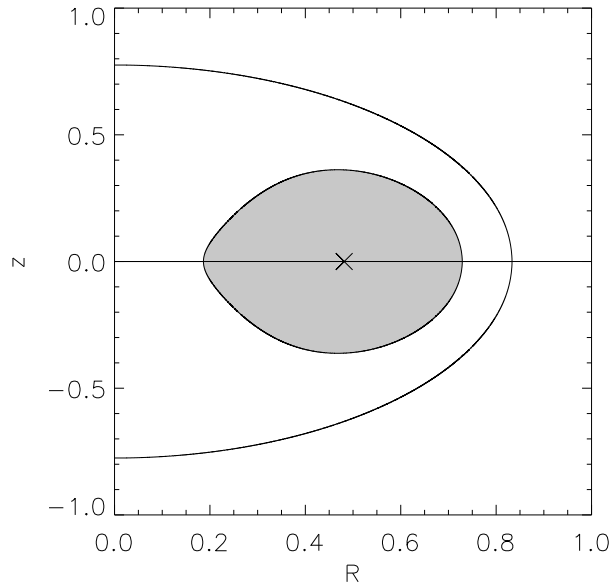,width=8.cm}} 
\caption{\label{fig:ZVC} The meridional-plane cross section of the
toroidal volume defined by the energy $E$ and angular momentum $L_z$
in an axisymmetric potential. Its boundary is the zero-velocity curve
ZVC. When $L_z=0$, the ZVC coincides with the equipotential (indicated
by the drawn curve), and the hole around the $z$-axis disappears. When
the angular momentum is maximal, the ZVC reduces to the circular orbit
(indicated by the cross).}
\end{figure}

We introduce the two-integral components (TICs) and calculate their
observable properties, and then describe how we use them to construct
two-integral models for galaxies. The properties of ICs follow by taking
$L_z=0$.

\subsection{Properties}

A TIC is completely specified by its energy $E_j$ and angular momentum
$L_{z,j}$ (where $j$ is a label for the TIC) and is defined by its
distribution function
\begin{eqnarray}
f_{[E_j,L_{z,j}]}=\!\left\{\begin{array}{l l}
     \! {\cal C}\,\delta(E\!-\!E_j)\,\delta(L_z\!-\!L_{z,j}) 
                                    &{\rm inside\,\,\,ZVC,}\\
     \! 0                           &{\rm elsewhere,}\end{array}\right.
\label{eq:tic-df}
\end{eqnarray}
with ${\cal C}$ a constant that is used to normalize the mass of the
TIC. The abbreviation ZVC denotes the zero-velocity curve in the $(R,
z)$-plane, defined as the locus of points for which
\begin{equation}
E_{\rm kin}=V_{\rm eff}(R,z)-E=0,
\label{eq:zvc-def}
\end{equation}
where 
\begin{equation}
V_{\rm eff} (R,z)=V(R,z)-\frac{L^2_z}{2\,R^2},
\label{eq:veff-def}
\end{equation}
is the effective potential. The ZVC defines a toroidal volume around
the $z$-axis (Fig.~\ref{fig:ZVC}) with inner and outer radii $R_1$ and
$R_2$.  The size of the central hole is determined by the value of the
angular momentum: when $L_z=0$, there is no hole at all ($R_1=0$) and
the torus fills the equipotential at the given energy completely, so that
the TIC reduces to the isotropic component (IC) at that energy. For the maximum 
value $L_z=R_c v_c$ (with $v^2_c=R_c\,\partial V_{\rm eff}(R_c)/\partial R_c$ 
the circular velocity at radius $R_c$) the torus reduces to the circular 
orbit at $E_j$, and $R_1=R_2=R_c$.

The density $\rho_j(R,z)$ of a TIC is defined as the integral of the
DF over all velocities, and is given by
\begin{equation}
\rho_j(R,z)=\left\{\begin{array}{l l}
{\!\displaystyle \frac{\pi\,{\cal C}}{R}}&{\rm inside\,\,\,ZVC,}\\
 \! 0               &{\rm elsewhere.}\end{array}\right.
\label{eq:tic-density}
\end{equation} 
The velocity moments are defined as the average of (powers of) the
velocity components over all velocities. From symmetry arguments and
from the fact that $f=f(E,L_z)$, the only nonzero first and second
moments are
\begin{eqnarray}
\langle v_\phi \rangle \!\! &=& \!\! 
            {\displaystyle \frac{\pi\,{\cal C}\,L_{z,j}}{R^2}}\nonumber\\
\langle v^2_\phi \rangle \!\! &=& \!\! 
          {\displaystyle \frac{\pi\,{\cal C}\,L^2_{z,j}}{R^3}}\nonumber\\
\langle v^2_R\rangle \!\! &=& \!\! \langle v^2_z\rangle = 
{\displaystyle {\pi\,{\cal C} \over R}} [V_{\rm eff}(R,z)-E_j], 
\label{eq:moments-def}
\end{eqnarray}
where these expressions are valid inside the ZVC. All moments vanish
outside the ZVC.

We assume that the galaxy is observed at an inclination $i$, so that
the projected coordinates are given by
\begin{eqnarray}
 x' \!\!&=&\!\! y\nonumber\\
 y' \!\!&=&\!\! -x\,\cos i + z\,\sin i\\
 z' \!\!&=&\!\! x\,\sin i + z\,\cos i,\nonumber
\label{eq:projection-def}
\end{eqnarray}
where $x'$ and $y'$ are chosen in the plane of the sky, with $y'$
along the projected minor axis of the galaxy, and $z'$ is measured
along the line of sight.

\begin{figure}
{\psfig{file=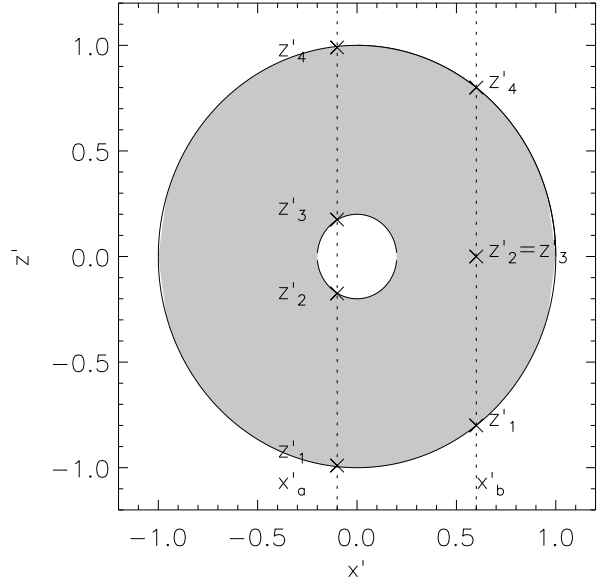,clip=,width=8.cm}} 
\caption{\label{fig:zvcproj} 
The line-of-sight crosses the ZVC in two (case b) or four (case a)
points, depending on the location of $(x',y')$. The plot shows an
edge-on configuration, so that the intersections are pairwize
symmetric.}
\end{figure}

The projected density $\Sigma_j(x',y')$ is the integral of the
intrinsic density along the line of sight
\begin{equation}
\Sigma_j(x',y')\!=\int\!\!\frac{\pi\,{\cal C}\,dz'}
                                {\sqrt{x'^2+(z'\sin i-y'\cos i)^2}}, 
\label{eq:surf-den-def}
\end{equation} 
which can be evaluated analytically. The intrinsic density of a TIC is
only defined for points that lie inside the toroidal volume defined by
the ZVC, so that the integral (\ref{eq:surf-den-def}) is defined on
the $z'$-intervals delimited by the two or four points where the
line-of-sight crosses the ZVC (Fig.~\ref{fig:zvcproj}). We give all
expressions for four intersections, labeled as $z'_1 < z'_2 \leq 0
\leq z'_3 < z'_4$. The case of two intersections follows by taking
$z'_2=z'_3=0$. We obtain
\begin{equation}
\Sigma_j(x',y')=\Sigma_{z'_4}-\Sigma_{z'_3}+\Sigma_{z'_2}-\Sigma_{z'_1}, 
\label{eq:surf-den-explicit}
\end{equation}
with
\begin{equation}
\Sigma_{z'_k}=\frac{\pi\,{\cal C}}{\sin i}\,
\ln\left[2\,\sin i \left(R'_{z'_k}+D_{z'_k}\right)\right], \qquad k=1,\ldots,4,
\label{eq:surfcart}
\end{equation}
where we have defined
\begin{eqnarray}
R'_{z'}=\sqrt{x'^2+(z'\sin i-y'\cos i)^2}
\label{eq:r-def}
\end{eqnarray}
and 
\begin{equation}
D_{z'}=z'\,\sin i-y'\,\cos i.
\label{eq:d-def}
\end{equation}
The limiting case of edge-on viewing ($i=0$) is given by
\begin{equation}
\Sigma_{z'_k}=\frac{\pi\,{\cal C}\,z'_k}{\sqrt{x'^2+y'^2}}, 
\qquad k=1,\ldots,4,
\label{eq:surfi0}
\end{equation}

The velocity profile $ {\cal L}_j(x',y',v_{z'})$, defined as the
distribution of stars over velocity $v_{z'}$ at the position
$(x',y')$, is the integral of $f(E,L_z)$ over the line of sight and
the two velocity components $v_{x'}$ and $v_{y'}$ in the plane of the
sky. Since the distribution function of a TIC is the product of two
$\delta$-functions, the triple integral in the expression for $ {\cal
L}_j$ reduces to a single integral,
\begin{equation}
{\cal L}_j(x',y',v_{z'})={\cal C}\int dz'\,J_{E_jL_{z,j}},
\label{eq:projVP}
\end{equation}
where the remaining integration in $z'$ is over the same intervals as
in eq.~(\ref{eq:surf-den-def}), and $J_{E_jL_{z,j}}$ is the Jacobian
for the change of variables from $(v_{x'},v_{y'})$ to $(E_j,L_{z,j})$.
It can be written as (C99)
\begin{equation}
J_{E_jL_{z,j}}=\frac{1}{\sqrt{A+B\,v_{z'}+C\,v^2_{z'}}},
\label{eq:jacobian}
\end{equation}
with 
\begin{eqnarray}
A\!\!&=&\!\! 2\left[V(x',y',z')\!-\!E_j\right](x'^2 \cos^2 i\!+\!D^2_{z'})\!-
 \!L^2_{z,j}, \nonumber\\
B\!\!&=&\!\! -2 x'\,L_{z,j}\,\sin i, \\
C\!\!&=&\!\! -R'^2_{z'}, \nonumber
\end{eqnarray}
with $R'_{z'}$ and $D_{z'}$ from eqs~(\ref{eq:r-def}) and (\ref{eq:d-def}), respectively.

The integral of the velocity profile over all allowed values of
$v_{z'}$ is equal to $\Sigma_j(x',y')$. The projected velocity moments
$\langle v^n_{z'} \rangle$ (with $n=1, 2, \ldots$) can be evaluated
similarly, resulting in
\begin{eqnarray}
\langle v_{z'}\rangle \!\!&=&\!\!-{\cal C}\,\int\!\!\frac{dz'}{R'^2_{z'}}
\left[J_{E_jL_{z,j}}-\frac{1}{2}\,B\,{\cal U}_{v_{z'}}\right]^{v_+}_{v_-}
\nonumber\\
\langle v^2_{z'}\rangle \!\!&=&\!\!-\frac{\cal C}{2}\,\int\!\!\frac{dz'}
{R'^2_{z'}}\left[(v-\frac{3B}{2C})J_{E_jL_{z,j}}+\right.\nonumber\\
&\,&\left.\frac{3\,B^2-4\,A\,C}{4\,C\,R'_{z'}}\,{\cal U}_{v_{z'}}\right]_{v_-}^{v_+},
\end{eqnarray}
with 
\begin{eqnarray}
{\cal U}_{v_{z'}}=\arcsin \bigg 
[\frac{v_{z'}\,R'_{z'}+x'L_{z,j}\sin i/R'_{z'}}{\sqrt{2\,E_{\rm kin}}
\sqrt{x'^2\cos^2 i\!+\!D^2_{z'}}}\bigg],
\label{eq:U}
\end{eqnarray}
and $J_{E_jL_{z,j}}$ is given in eq.~(\ref{eq:jacobian}). The values 
$v_-$ and $v_+$ are the limits of the integration in $v_{z'}$. 

\subsection{Schwarzschild's method}
\label{sec:schwarzschild}

The individual TICs can be used to model galaxies with Schwarzschild's
orbit superposition method. The procedure follows the usual
steps (e.g., C99). First determine the density distribution, either
from a theoretical model or by deprojecting the observed projected
surface density of a galaxy and choosing a mass-to-light ratio. Then
calculate the corresponding gravitational potential from Poisson's
equation. In this potential, define a collection of TICs by specifying
the integrals $(E_j, L_{z,j})$.  Jeans' theorem guarantees that a
sampling of the full range in energies and angular momenta results in
a TIC library that is representative for this potential. This can be
achieved by sampling the radius on a grid of $n_E$ circular radii
$\{R_c\}$, chosen e.g. logarithmically on a range that includes $\geq
99.9$\% of the mass. From this, the energy grid $\{E_c\}$ follows by
calculating the energy of the circular orbit at the $\{R_c\}$. The
$n_{L_z}$ angular momenta can then be chosen linearly between the
minimum ($L_z=0$, orbits confined to the meridional plane) and maximum
($L_z=R_c v_c$, the circular orbit) at every $E_c$.

This choice of grids defines a library of $n_t = n_E\, n_{L_z}$
TICs. Schwarzschild's method then determines the weighted sum of 
the TICs in this library that is closest to a data set $P_i$ (generally 
consisting of both photometric and kinematic data). If $O_{ij}$ is the 
contribution of the $j$-th TIC to the $i$-th aperture, the problem 
reduces to solving for the weights $\gamma_j$ in
\begin{equation}
\sum_j^{N_t}\gamma_j\,O_{ij} = P_i, \qquad i=1,\ldots,N_p,
\label{eq:matrix}
\end{equation}
where $N_p$ is the number of apertures for which observations are available.
The $\gamma_j$, which determine the weight of each individual 
TIC in this superposition, are found by using a least-squares method, 
with the additional constraint that $\gamma_j\geq 0$ (e.g., R97). Furthermore,
$\gamma=\gamma(E,L_z)$, since every TIC is labeled by a unique
combination of $E$ and $L_z$. The DF that corresponds to this set of
weights can then be found from (Vandervoort 1984)
\begin{equation}
{\rm DF}(E,L_z)=\frac{\gamma(E,L_z)}{m_{E,L_z}\,A_{E,L_z}},
\label{eq:gamdf}
\end{equation}
where $m_{E,L_z}$ is the mass of the TIC and $A_{E,L_z}$ is the area
of the cell around $E$ and $L_z$ in integral space.

Since the matrix problem of eq.~(\ref{eq:matrix}) is often
ill-conditioned, the weights and therefore the reconstructed DF may vary
rapidly as a function of $E$ and $L_z$. This can be avoided by adding 
regularisation constraints to the problem (e.g., Zhao 1996; R97). These 
constraints force the weights $\gamma(E,L_z)$ towards a smooth function 
of $E$ and $L_z$ by minimizing the $n$-th order derivatives
$\partial^n\gamma(E,L_z)/\partial E^n,\partial^n\gamma(E,L_z)/\partial
L^n_z$.  The degree of smoothening is determined by the order $n$ and
by the maximum value that the derivatives are allowed to have.

The predicted observables that are obtained by calculating the 
weighted sum of the TICs usually differ from the observed values $P_i$. 
There are various reasons for this: the DF of the galaxy may not be of 
the form $f(E,L_z)$, model parameters such as inclination, stellar 
mass-to-light ratio or central black hole mass may not be chosen properly, 
the observations can contain systematical errors, and the calculations are 
done on finite grids, which introduces discretization effects. To measure 
the discrepancy, we define a quality-of-fit parameter, $\chi^2$, as
\begin{equation}
\chi^{2}=\sum_{i=1}^{N_p}\left(\frac{P^*_i-P_i}{\Delta P_i}\right)^{2},
\label{eq:chisq}
\end{equation}
in which $P^*_i$ is the model prediction at the $i$-th aperture and
$\Delta P_i$ is the (observational) uncertainty that is associated 
with $P_i$.

\section{Numerical Implementation}
\label{sec:implementation}

The numerical implementation of Schwarzschild's method for
three-integral axisymmetric models including kinematic constraints is
described in detail in C99. Here we discuss the simplifications for
models containing TICs, which can be used to speed up the
calculations. Readers who are more interested in applications, can
skip to \S\ref{sec:examples}.

\subsection{Storage grids}
\label{sec:general}

All observables are calculated on grids that are adapted to the size
of the galaxy and to the constraints that have to be reproduced. The
meridional plane density is stored on a polar grid in the meridional
plane, with a radial grid that is sampled logarithmically on the same
range as the $\{R_c\}$ and a linear grid in polar angle between $0$
and $\pi/2$. The projected density is calculated on a Cartesian grid
in the sky plane, and the velocity profile is stored on a
three-dimensional Cartesian grid in $x',y'$ and $v_{z'}$. The bin
sizes of both grids are adapted to the resolution of the kinematic
data to be reproduced. The observables are averaged over these grids,
so that a continuous observable $O_j(\vec{x},\vec{v})$ is represented
by a discrete set of averaged grid entries 
\begin{equation}
\label{eq:anorm}
\langle O_j \rangle=\int_{\rm cell} O_j(\vec{x},\vec{v}).
\end{equation}
All observables are related to the DF through triple (intrinsic density 
and velocity profile) or quadruple (projected density) integrals, two of 
which are trivial for TICs, because of their $\delta$-function behaviour.
This means that the averages $\langle O_j \rangle$ can be written
as no worse than quadruple integrals over the region enclosed by the
cross-section of the ZVC and the grid cell. Depending on the relative
position of the grid cell and the ZVC, the four integrals can be
interchanged and the expressions can be simplified. In the following,
we show how this is achieved for all observables.

\begin{figure}
{\psfig{file=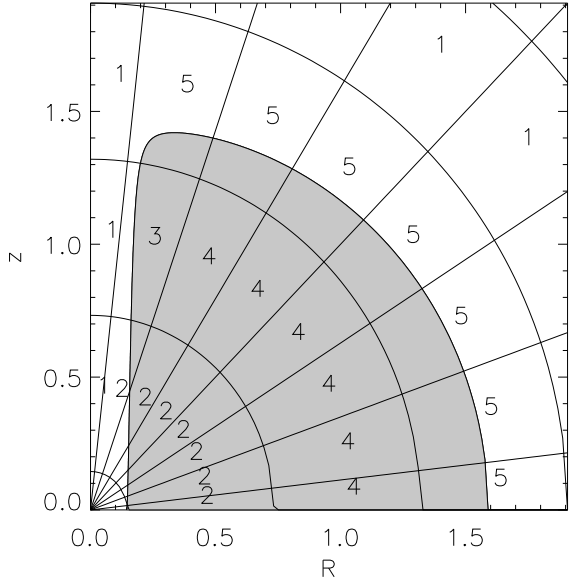,width=8.cm}} 
\caption{\label{fig:ZVCgrid} A ZVC$(E_j,L_{z,j})$ in the meridional
 plane (the grey region), together with the cells of the meridional
 plane grid in radius and angle. The labels denote the method of
 integration (ana\-ly\-tical or numerical, see text). }
\end{figure}

\subsection{Meridional plane density}
\label{sec:mermass}

The simple form of the meridional plane density (\ref{eq:tic-density})
leads to considerable simplifications in the evaluation of the
averaged meridional plane density $\langle \rho_j \rangle$. The
contribution of a general grid cell bordered by
$(r_-,r_+)$ and $(\theta_-,\theta_+)$ is
\begin{equation}
\label{eq:numerical}
\langle \rho_j\rangle
=2\pi^2\,{\cal C}\int\limits_{r_-}^{r_+}\!dr 
      \int\limits_{\theta_-}^{\theta_+} \!d\theta\,\frac{r^2\,\sin\theta}{R}.
\end{equation}
For cells that are completely surrounded by the ZVC, which are
relatively numerous for TICs with small energies (corresponding to
large ZVCs), this integral can be evaluated and is given by
\begin{eqnarray}
\label{eq:analytical}
\langle \rho_j \rangle
=\pi^2\,C\left(r^2_+-r^2_-\right)\left(\theta_+-\theta_-\right).
\end{eqnarray}
When, on the other hand, $E_j$ is large, so that the ZVC is less
extended, the cell is (much) larger than the ZVC. This means that the
integration region in eq.~(\ref{eq:numerical}) can be adjusted to
match the ZVC size better, which speeds up the calculations
considerably. Table~\ref{tab:table1} and Fig.~\ref{fig:ZVCgrid} show
for which cells one can apply these two simplifications. Calculating
all meridional plane densities for a library of 70 x 20 TICs takes a few
minutes on a machine with a 1 GHz processor.

\begin{center}
\begin{table}
\begin{tabular}{l  c  c c c c}
\hline\hline
Cell&$r_- \in$& $r_+ \in$&$\theta_- \in$&Eq.&New range\\\hline
$1$&$[0,R_1]$&$[0,R_1]$&-&$0$   &-\\
$2$&$[0,R_1]$&$[R_1,R_2]$&-&$\ref{eq:numerical}$& $[R_1,r_+]$\\
$*$&$[0,R_1]$&$[R_2,\infty\rangle$ &-& $\ref{eq:numerical}$&$[R_1,R_2]$\\
$3$&$[R_1,R_2]$&$[R_1,R_2]$&$[0,\theta_{\rm ZVC}]$
                                      & $\ref{eq:numerical}$&$[r_-,r_+]$\\
$4$&$[R_1,R_2]$&$[R_1,R_2]$&$[\theta_{\rm ZVC},\frac{1}{2}\pi]$
                                           &$\ref{eq:analytical}$&$-$\\
$5$ & $[R_1,R_2]$&$[R_2,\infty \rangle$ &-
                                  &$\ref{eq:numerical}$& $[r_-,R_2]$\\\hline
\end{tabular}
\caption{\label{tab:table1} 
For each cell in Fig.~\ref{fig:ZVCgrid} (and for one cell that is not
shown, labeled with $*$), the simplifications in the calculation of
the averaged meridional plane density are given. The equation that has
to be used is listed, or a value of $0$ when the cell is located
outside the ZVC. When eq.~(\ref{eq:numerical}) applies, the new
integration limits are given.}
\end{table}
\end{center}

\subsection{Projected density}

The average of the projected density (\ref{eq:surfcart}) over a cell in a
Cartesian grid on the sky, bordered by $(x_-,x_+)$ and $(y_-,y_+)$, is
given by (cf.\ eqs~\ref{eq:surf-den-explicit} and \ref{eq:surfcart})
\begin{equation}
\langle\Sigma_j\rangle=\langle\Sigma_{z'_4}\rangle-\langle\Sigma_{z'_3}\rangle+
\langle\Sigma_{z'_2}\rangle-\langle\Sigma_{z'_1}\rangle, 
\end{equation}
with
\begin{equation}
\label{eq:Sigma}
\langle \Sigma_{z'_k}\rangle
=\frac{\pi\,{\cal C}}{\sin i}\,\int\limits_{x'_-}^{x'_+}
 \!dx'\int\limits_{y'_-}^{y'_+}\!dy'\,
\ln\left[2\,\sin i \left(R'_{z'_k}+D_{z'_k}\right)\right],
\end{equation}
(or, when $i=0$, the integral over eq.~(\ref{eq:surfi0})).
Since $z'_k=z'_k(x',y')$, this expression cannot be simplified further
without knowledge of the relation between $z'_k$ and $x',y'$.
Unfortunately, this relation is usually complicated and results in a
very lengthy integrand, which rules out analytical evaluation.
Therefore, we integrate eq.~(\ref{eq:Sigma}) numerically by using a
routine for more-dimensional numerical quadrature from the NAG-library,
{\sc D01FCF}. These calculations are slightly more time-consuming 
than the ones for the meridional plane density: calculating the 
averaged projected densities for a library of 1400 TICs on the 
machine mentioned in \S\ref{sec:mermass} takes of the order of thirty minutes.

\begin{figure}
{\psfig{file=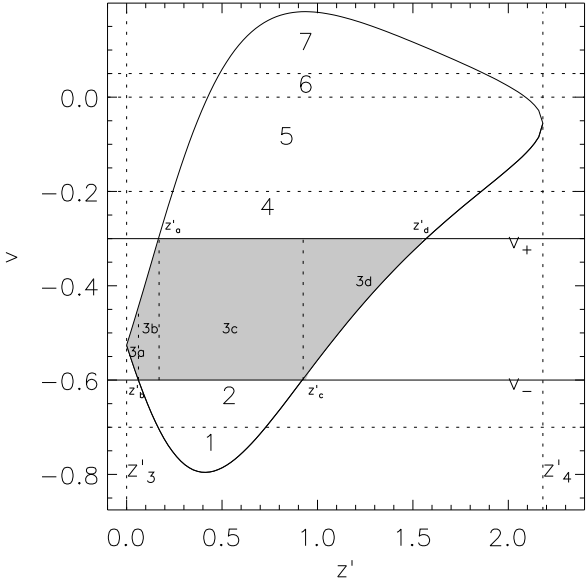,width=8.truecm,clip=}} 
\caption{\label{fig:manta} The extreme velocities $v_{1,2}$ of the
velocity profile as a function of the line-of-sight coordinate $z'$,
at fixed and nonzero $x'$ and $y'$. Only positive values of $z'$ are drawn,
a similar configuration exists for $z'<0$. The lower curve is $v_1$, 
and the upper curve is $v_2$. The horizontal dotted lines illustrate 
possible locations of the velocity bin ($v_-,v_+$). The domain of 
integration in $v$ and $z'$ is the region enclosed by the curve and 
these bins (the grey area is an example; the new integration limits, 
$z'_a,z'_b, z'_c,z'_d$ are indicated).}
\end{figure}\looseness=-2

\subsection{Velocity Profile}

The averaged ${\cal L}$ in a cell bordered by $(x_-,x_+)$, $(y_-,y_+)$
and $(v_-,v_+)$ equals
\begin{equation}
\label{eq:vp_one}
\langle {\cal L}\rangle={\cal C}\int\limits_{x'_-}^{x'_+}\!dx'
 \int\limits_{y'_-}^{y'_+}\!dy' \int\limits_{v_-}^{v_+}\!dv_{z'} \!\!
\int\limits_{z'(v)_-}^{z'(v)_+} dz'\,J_{E_j,L_{z,j}}.
\end{equation}
This integral is only defined on the region where the Jacobian exists,
which is bordered by the points for which $A+B\,v_{z'}+C\,v^2_{z'} =
0$ (\ref{eq:jacobian}). We solve this equation for $v_{z'}$ to find
the following relation between the minimum and maximum velocities of
the velocity profile, $v_1,v_2$, and $z'$:
\begin{equation}
\label{eq:extremev}
v_{1,2}(z')\!=\!\frac{-x'\,L_{z,j}\sin i \pm 
 R'_{z'}\sqrt{2\,E_{\rm kin}}\sqrt{x'^2\cos^2 i\!+\!D^2_{z'}}}{R'^2_{z'}},
\end{equation}
with $R'_{z'}$ and $D_{z'}$ given in eqs~(\ref{eq:r-def})
and (\ref{eq:d-def}), respectively. The solid curve in
Fig.~\ref{fig:manta} shows these extreme velocities as a function of
$z'$, for arbitrary $x'$ and $y'$. From eq.~(\ref{eq:extremev}), the 
$v_{1,2}$ are directly proportional to $E_{\rm kin}$, which
means they only exist inside the intervals $[z'_1,z'_2]$ and
$[z'_3,z'_4]$. When $z'=z'_i$, $E_{\rm kin}=0$, and the cut-off
velocities are
\begin{equation}
\hat{v}_{1,2}(z')=-\frac{x'\,L_{z,j}\,\sin i}{R'^2_{z'}},
\end{equation} 
where $\hat{v}_1$ is the velocity at $z'=z'_2,z'_3$ and $\hat{v}_2$ is
reached when $z'=z'_1,z'_4$. It follows that $v_1=-v_2$ at $x'=0$, so
that ${\cal L} $ is symmetric around $v=0$.

\begin{figure}
{\psfig{file=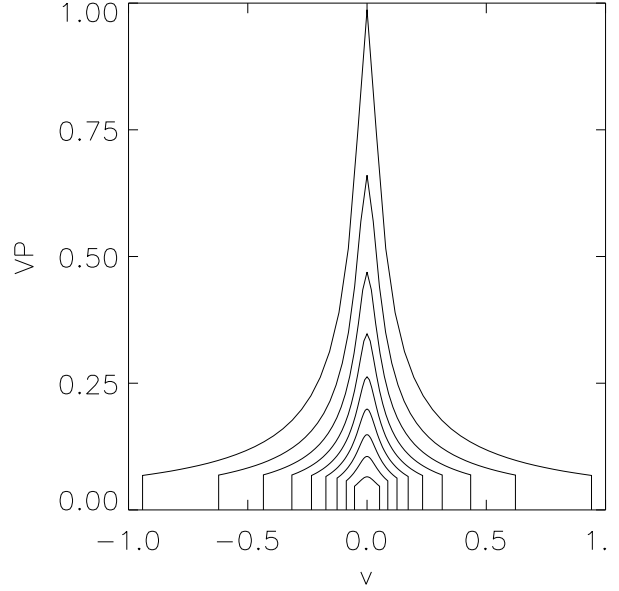,width=8.truecm,clip=}} 
\caption{\label{fig:VPexamples} A series of velocity profiles along
the minor axis, for a double power law density distribution 
(\ref{eq:double_power_law}). The formalism that is outlined in the text
is used to calculate ${\cal L}$, but no average is made over the grid
cells in $x'$ and $y'$ and very narrow bins in velocity are used to
ensure that the resulting curve is smooth. }
\end{figure}\looseness=-2

Since $v$ is integrated over the interval $v_-,v_+$, the limits of
integration in the inner integral of eq.~(\ref{eq:vp_one}),
$z'(v)_-,z'(v)_+$, are determined by the cross-section of the velocity
bin and the region enclosed by $v_1$ and $v_2$ (the grey area in
Fig.~\ref{fig:manta} is an example).  The $z'$- integration cannot in
general be carried out analytically. However, we can interchange the
inner two integrals and, by using the quadratic dependence of the
Jacobian on $v_{z'}$, evaluate the integration over $v_{z'}$:
\begin{eqnarray}
\label{eq:numerical2}
\langle{\cal L}\rangle_{z'_-,z'_+,v_-,v_+}={\cal C}
 \int\limits_{x'_-}^{x'_+}\!\!dx'\int\limits_{y'_-}^{y'_+}\!\!dy' 
 \int\limits^{z'_+}_{z'_-}\!\! dz'\,\frac{({\cal U}_{v_+}-{\cal U}_{v_-})}{R'_{z'}},
\end{eqnarray}
where we have used eq. (\ref{eq:U}). This expression can be simplified even 
further, since for the values of $z'$ where the bin extends outside the region 
enclosed by $v_1$ and $v_2$ ($v_+ > v_2$, $v_- < v_1$), we can replace 
$v_{-,+}$ with $v_{1,2}$. In this case, we substitute eq.~(\ref{eq:extremev}) 
in eq.~(\ref{eq:numerical2}), carry out the integration over $z'$ and use
eq.~(\ref{eq:Sigma}). The averaged velocity profile then equals
\begin{equation}
\label{eq:analytical2}
\langle {\cal L}\rangle_{z'_-,z'_+,v_1,v_2}=
 \langle\Sigma_{z'_+}\rangle-\langle\Sigma_{z'_-}\rangle.
\end{equation}
By calculating the intersections (if any) of $v_-$ and $v_+$ with
$v_{1,2}$ (respectively, $z'_b < z'_c$ and $z'_a < z'_d$), we separate
the integration interval in $z'$ in regions where we use eq.~(\ref{eq:analytical2}) 
and ot\-hers where eq.~(\ref{eq:numerical2}) applies. The averaged velocity 
profile in the grey area of Fig. ~\ref{fig:manta} therefore consists of 
the averaged ${\cal L}$ in the areas $\{z'_3,\,z'_b\} \times 
\{v_1(z'),\,v_2(z')\}$ (area 3a), $\{z'_b,\,z'_a\} \times \{v_-,\,v_2(z')\}$ 
(area 3b), $\{z'_a,\,z'_c\} \times \{v_-,\,v_+\}$ (area 3c), and 
$\{z'_c,\,z'_d\} \times \{v_1(z'),\,v_+\}$ (area 3d), which contribute, 
respectively, $\langle {\cal L}\rangle_{z'_3,z'_b,v_1,v_2}$,
$\langle{\cal L}\rangle_{z'_b,z'_a,v_-,v_2}$, $\langle {\cal L}
\rangle_{z'_a,z'_c,v_-,v_+}$ and $\langle {\cal L}\rangle_{z'_c,z'_d,v_1,v_+}$.
The total averaged velocity profile in this cell is then given by the sum of these four
terms. A similar reasoning applies to all other bins in Fig.~\ref{fig:manta} 
(Table~\ref{tab:table2}). To check whether the resulting ${\cal L}$ is
smooth and resembles those calculated by C99, we have calculated a
series of ${\cal L}$'s along the minor axis (which means they are
symmetric around $v=0$). The result is shown in Fig.~\ref{fig:VPexamples}, 
which can be compared to, e.g., Fig. 5 of C99.\looseness=-2

\begin{table}
\begin{center}
\begin{tabular}{l l l l}
\hline\hline
Cell&$v_- \in$& $v_+ \in$& $z',v_{z'} \in$\\\hline
$1$&$\langle-\infty,\hat{v}_1]$&$\langle-\infty,\hat{v}_1]$&$\{z'_a,z'_d\}\times\{v_1,v_+\}$\\
$2$&$\langle-\infty,\hat{v}_1]$&$\langle-\infty,\hat{v}_1]$&$\{z'_a,z'_b\}\times\{v_1,v_+\}$\\
   &                           &                             &$\{z'_b,z'_c\}\times\{v_-,v_+\}$\\
   &                           &                           &$\{z'_c,z'_d\}\times\{v_1,v_+\}$\\
$3$&$\langle-\infty,\hat{v}_1]$&$[\hat{v}_1,\hat{v}_2]$&$\{z'_3,z'_b\}\times\{v_1,v_2\}$\\
   &                           &                       &$\{z'_b,z'_a\}\times\{v_-,v_2\}$\\
   &                           &                       &$\{z'_a,z'_c\}\times\{v_-,v_+\}$\\
   &                           &                       &$\{z'_c,z'_d\}\times\{v_1,v_+\}$\\
$*$&$\langle-\infty,\hat{v}_1]$&$[\hat{v}_2,\infty\rangle$&$\{z'_3,z'_4\}\times\{v_1,v_2\}$\\
$4$&$[\hat{v}_1,\hat{v}_2]$&$[\hat{v}_1,\hat{v}_2]$&$\{z'_b,z'_a\}\times\{v_-,v_2\}$\\
   &                       &                       &$\{z'_a,z'_c\}\times\{v_-,v_+\}$\\
   &                       &                       &$\{z'_c,z'_d\}\times\{v_1,v_+\}$\\
$5$&$[\hat{v}_1,\hat{v}_2]$&$[\hat{v}_2,\infty\rangle$&$\{z'_d,z'_b\}\times\{v_1,v_2\}$\\
   &                       &                          &$\{z'_b,z'_a\}\times\{v_-,v_2\}$\\
   &                       &                          &$\{z'_a,z'_d\}\times\{v_-,v_+\}$\\
   &                       &                          &$\{z'_d,z'_b\}\times\{v_1,v_+\}$\\
$6$&$[\hat{v}_2,\infty\rangle$&$[\hat{v}_2,\infty\rangle$&$\{z'_b,z'_a\}\times\{v_-,v_2\}$\\
   &                          &                          &$\{z'_a,z'_b\}\times\{v_-,v_+\}$\\
   &                          &                          &$\{z'_b,z'_d\}\times\{v_-,v_2\}$\\
$7$&$[\hat{v}_2,\infty\rangle$&$[\hat{v}_2,\infty\rangle$&$\{z'_a,z'_d\}\times\{v_-,v_2\}$
\\\hline
\end{tabular}
\end{center}
\caption{\label{tab:table2} The first column gives the label of the
velocity bins in Fig.~\ref{fig:manta} (and for one bin that is not
shown, denoted with $*$). The second and third column give the
intervals in which the limits of the velocity bin, $v_-$ and $v_+$,
are located. In the fourth column, the domains in ($z',v_{z'}$) on 
which eq.~(\ref{eq:numerical2}) and eq.~(\ref{eq:analytical2}) are
defined, are given. The total velocity profile is then obtained by 
adding the contributions of all these regions.}
\end{table}

\subsection{PSF convolution}

In the previous subsections, we described how to obtain the average of
the TIC observables over the cells of the storage grids. These quantities 
can only be compared with observational data when point spread function
(PSF) effects are taken into account correctly. This means that 
observationally accessible quantities such as the projected density and 
the velocity profile have to be convolved with the appropriate PSF, 
which is usually described in terms of a (sum of) Gaussian functions. The 
extra two-dimensional integration that this convolution introduces can be 
carried out analytically when it is combined with the average over the 
storage cell (see Appendix D of Q95). This means that an extra term, 
consisting of error functions, appears in the expressions for the observables 
that were given previously.

\vskip 1.0truecm

\section{Examples and tests}
\label{sec:examples}

To test our method, we apply the numerical implementation to three
theoretical potential-density pairs. We investigate whether our
formalism rules out the same models as previous methods and determine
the accuracy with which a theoretical distribution function can be
reproduced. This also allows us to study the behaviour of the $\chi^2$
(\ref{eq:chisq}) and the regularisation algorithm, which is
similar to the one that is used in the three-integral Schwarzschild
methods (C99). Applications to two-dimensional photometry and
kinematics of galaxies are described in Verolme et al. and
McDermid et al.

\begin{figure}
{\psfig{file=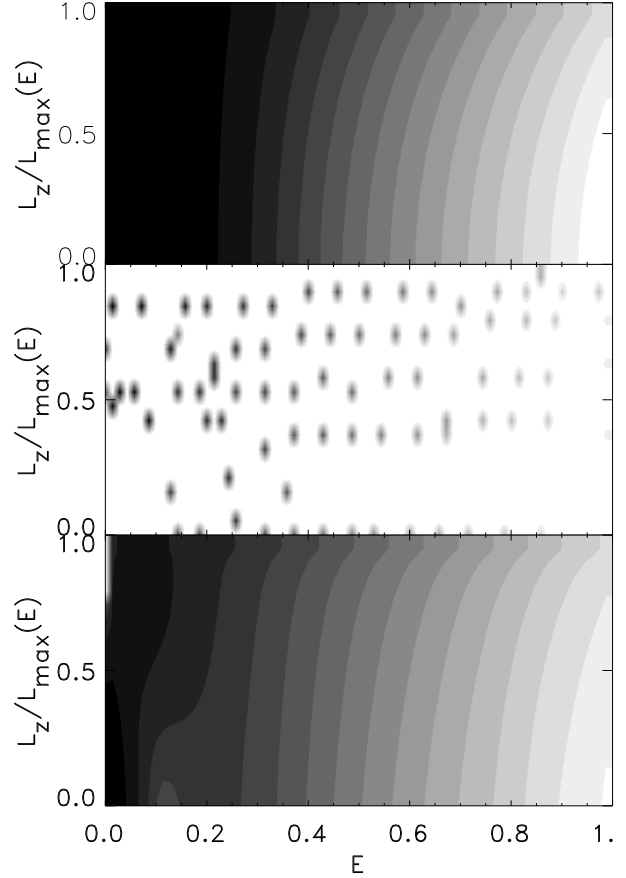,clip=,width=8.cm}} 
\caption{\label{fig:DFs} The upper panel shows the Kuzmin-Kutuzov DF
(\ref{eq:DFkk}) as a function of energy $E$ and fractional
angular momentum $L_z/L_{z,{\rm max}}$. The contour levels are chosen
logarithmically and darker shading corresponds to larger values of 
the DF. The reconstructed DF (\ref{eq:gamdf})
for a two-integral Schwarzschild model that fits only the meridional
plane masses is shown in the middle panel. There is a large
discrepancy between the two functions. The lower panel shows the smooth
DF that is found by including regularisation constraints in the
fit. The contours are drawn at the same levels as in the top panel,
showing that this DF reproduces the theoretical model very
closely. The fractional difference between the two DFs on a range
that includes $>99\%$ of the mass is smaller than $10^{-3}$.}
\end{figure}

\subsection{Regularisation: Kuzmin-Kutuzov model}
\label{sec:kk}
We first test the ability of our method to reproduce a known
$f(E,L_z)$, given $\rho$ and $V$, and investigate the effect of
regularisation on the results. We use the density and potential of the
Kuzmin--Kutuzov (1962) model
\begin{eqnarray}
\label{eq:rhoKK}
\rho(R,z) \!\!\!\!\!&=&\!\!\!\!\!\frac{M\,c^2}{4\,\pi}
\frac{(a^2+c^2)\,R^2+2\,a^2\,(z^2+c^2)+a^4+3\,a^2\,u}
{u^3\,(R^2+z^2+a^2+c^2+2\,u)^3},\nonumber\\
V(R,z)\!\!\!\!\!&=&\!\!\!\!\!\frac{G\,M}{\sqrt{R^2+z^2+a^2+c^2+2\,u}},
\end{eqnarray}
with $u=\sqrt{c^2\,R^2+a^2\,(z^2+c^2)}$. The density distribution is
nearly spheroidal, and many of the properties of these models,
including the projected surface density, can be evaluated
explicitly. 

The even part of the $f(E, L_z)$ distribution function for the 
Kuzmin--Kutuzov models is given by (Dejonghe \& de Zeeuw 1988; Batsleer 
\& Dejonghe 1993)
\begin{eqnarray}
\label{eq:DFkk}
&\!\!\,\!\!\!&f(E,L_z)=\frac{c^2 E^{5/2}}{\sqrt{8}a\pi^3}
                                         \sum\limits_{\epsilon=-1,1}\\
&\!\!\!\,\!\!\!&\int\limits_{0}^{1}\!\!\!\,dt
\frac{(1-t^2)\,\left[(3+4\,x_\epsilon-x^2_\epsilon)(1-x_\epsilon)(1-t^2)+
12\,t^2\right]}{(1-2\,a\,E\,t\,\sqrt{1-t^2}
                               +\epsilon\,L_z\,t\sqrt{2\,h\,E})^5},\nonumber
\end{eqnarray}
with 
\begin{eqnarray}
x_\epsilon(t)=2\,a\,E\,t\,\frac{\sqrt{1-t^2}}{1+\epsilon\,\sqrt{z}\,t}.
\end{eqnarray}
This integral can be expressed in terms of elementary functions, but
this leads to an exceedingly lengthy expression, and it is best
evaluated numerically. A grey-scale plot of the Kuzmin-Kutuzov DF as a
function of energy and fractional angular momentum is shown in the
upper panel of Fig.~\ref{fig:DFs}.

We attempted to reproduce this DF by calculating a
two-integral Schwarzschild model that fits the meridional plane mass
(\ref{eq:rhoKK}). We used $n_E=70$ and $n_{L_z}=20$, including $L_z=0$, 
and adopted $16$ radial and $7$ angular bins, like C99. With these 
parameters, the model density was fitted with very high accuracy 
(fractional difference smaller than $0.001$), but the TIC weights vary 
dramatically with energy and angular momentum, resulting in a very 
jagged DF (the middle panel in Fig.~\ref{fig:DFs}). Including regularisation 
constraints enforces the weights to vary smoothly with TIC index, resulting 
in a DF that reproduces the theoretical DF to within 0.1\%. This is 
illustrated in the lower panel of Fig.~\ref{fig:DFs}, which was obtained 
by requiring the same fitting accuracy ($\Delta P$ in eq.~\ref{eq:chisq}) 
for both the meridional plane masses and the regularisation constraints. 
This means that both constraint types have equal influence on the final
model, so that the model DF is smooth, but still reproduces the density.

\begin{figure}
{\psfig{file=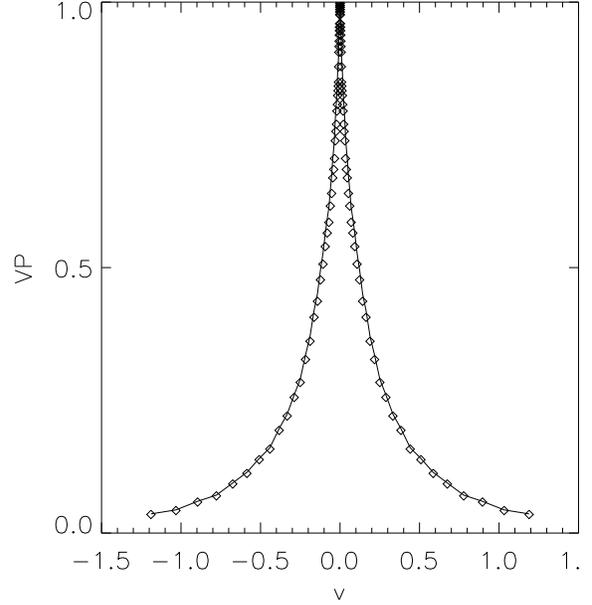,clip=,width=8.cm}} 
\caption{\label{fig:VPcomparison} The theoretical velocity profile for a
point on the minor axis, calculated directly from eq.~(\ref{eq:gamdf}) 
(the solid line), and the superposition of all individual TIC velocity 
profiles (indicated with diamonds) for the Kuzmin-Kutuzov density 
distribution of eq.~(\ref{eq:rhoKK}). The two curves agree to $0.1\%$.}
\end{figure}

In order to measure how fitting to the mass determines the
kinematics of the model, we have calculated a theoretical velocity
profile on the minor axis directly from the distribution function 
eq.~(\ref{eq:gamdf}), and compared this with the predicted profile, 
given by the superposition of all the individual TIC velocity profiles. 
Fig.~\ref{fig:VPcomparison} demonstrates that
the agreement between the two curves is significantly better than the
typical measurement error in observed velocity profiles (van der Marel 
\& Franx 1993; Joseph et al. 2001).  

\subsection{Models with a central black hole}
\label{sec:bh}

We now consider $f(E,L_z)$ models with a central black hole,
represented by a point mass with potential $GM_\bullet /r$.  

For spherically symmetric galaxies, there is a direct relation between
the luminous density distribution and the isotropic distribution
function $f(E)$ (Eddington 1916).  When the DF is required to be
non-negative, this relation implies that no spherical, isotropic
galaxies with a central black hole exist for cusp slopes shallower
than $-0.5$ (e.g., Binney \& Tremaine 1987). An indication that this
restriction disappears when a higher degree of anisotropy is allowed
is illustrated by the $f(E,L^2)$ (with $L$ the angular momentum) that
is obtained by placing all stars on circular orbits (Hunter 1975a).

To investigate this effect for general anisotropic distribution
functions $f(E,L_z)$, Q95 applied the HQ-method to density
distributions that are stratified on similar concentric spheroids.
They considered models with a density profile that follows a double
power law, given by
\begin{eqnarray}
 \label{eq:double_power_law} 
 \rho = \rho_{0}\left(\frac{m}{b}\right)^{\alpha}\left[1 + \left
 (\frac{m}{b}\right)^{2}\right]^{\beta},
\end{eqnarray}
where $\rho_0$ and $\alpha$ are the central density and cusp slope,
$m^2=R^2+z^2/q^2$ with $q$ the intrinsic axis ratio, and $b$ is 
the break radius. Q95 found that all oblate ($q<1$) double power law
profiles correspond to a non-negative distribution function. However,
adding a central black to these models limits the central 
cusp slope to $\alpha < -0.5$, as in the spherical case. This restriction is
independent of black hole mass, as long as $M_{\bullet}> 0$. 

De Bruijne et al. (1996) showed that this result is valid for a much
larger family of cusped spheroidal density profiles with a central
black hole: when the potential is spherically symmetric, the
distribution function
\begin{eqnarray}
\label{dfjos}
f(E,L^2,L_z)=\sum\limits_{\mu,\lambda}\left(\frac{L}{L_{\rm max}}\right)^
{\mu}\,\left(\frac{L_z}{L_{\rm max}}\right)^{\lambda}\,g(E),
\end{eqnarray}
corresponds to the density
\begin{eqnarray}
\label{densjos}
\rho(r,\theta)=\sum\limits_{\mu,\lambda}D^{\mu,\lambda}r^{-\alpha}
\sin^{2\lambda}\theta,
\end{eqnarray}
in which the $D^{\mu,\lambda}$ are given in terms of $\Gamma$-functions 
by eq. (17) of de Bruijne et al. The argument of a $\Gamma$-function
must be positive, which implies that, near a black hole, where the potential 
falls off as $1/r$ ($\delta=1$ in their notation), eqs~(\ref{dfjos}) and 
(\ref{densjos}) exist only for
\begin{eqnarray}
\label{condition1}
\alpha < \lambda-\mu-{\textstyle \frac12}.
\end{eqnarray}
We see from eq.~(\ref{dfjos}) that the limiting case of a two-integral
DF $f(E,L_z)$ is given by $\mu=0$, so that eq.~(\ref{condition1})
reduces to
\begin{eqnarray}
\label{condition2}
\alpha < \lambda-{\textstyle \frac12}.
\end{eqnarray}
This means that density distributions that can be expanded in terms of
$\sin \theta$ must obey condition (\ref{condition2}) on each of the
terms of the expansion in order to have a physical DF near the black
hole. The lowest-order term of this expansion, usually with
$\lambda=0$, places the most stringent condition on $\alpha$,
i.e. $\alpha < -0.5$.

These results show that analytical methods are very powerful for
deriving criteria for the existence of models. In contrast,
Schwarzschild's method always finds an orbit superposition that fits
the constraints to some degree. The $\chi^2$ (\ref{eq:chisq}),
determined from residual errors, is the only measure for the existence
of the underlying distribution function. A priori, it is not clear
whether this parameter is able to reproduce the sharp limits on the
model parameters that are found with analytical methods. For this
reason, we constructed two-integral models for the density profile 
(\ref{eq:double_power_law}). We added a central point mass
$M_{\bullet}\sim 1$\% of the total galaxy mass, mimicking a central
black hole, and varied the cusp slope at constant flattening.  As in
\S\ref{sec:kk}, we took $n_E=70$ and $n_{L_z}=20$ (again including $L_z=0$), 
and adopted 16 radial and 7 angular bins.\looseness=-2

\begin{figure}
{\psfig{file=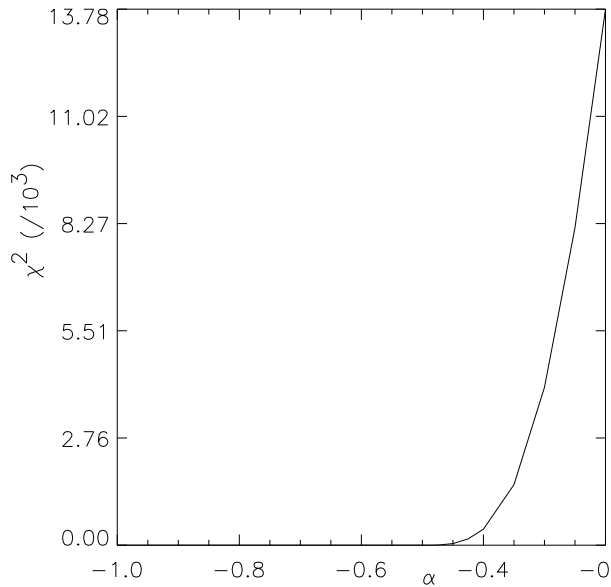,clip=,width=8.cm}} 
\caption{\label{fig:twoint} The quality-of-fit $\chi^2$ for selfconsistent
 two-integral models as a function of cusp slope $\alpha$, at constant
 intrinsic flattening. The $\chi^2$ is large for values in the
 interval $[-0.5,0]$, and negligible for $\alpha < -0.5$. This is 
 caused by the non-existence of the self-consistent $f(E, L_z)$ models 
 with $\alpha > -0.5$. }
\end{figure}

Fig.~\ref{fig:twoint} shows the resulting $\chi^2$ as function of
$\alpha$ for self-consistent two-integral models. While $\chi^2 <
10^{-5}$ for values of $\alpha$ below $-0.5$, it increases rapidly as
soon as $\alpha$ increases above $-0.5$. This shows that our method
recovers the models of Q95, and that our goodness-of-fit parameter
$\chi^2$ is a good discriminator of the (non)-existence of a given
model.

\begin{figure}
{\psfig{file=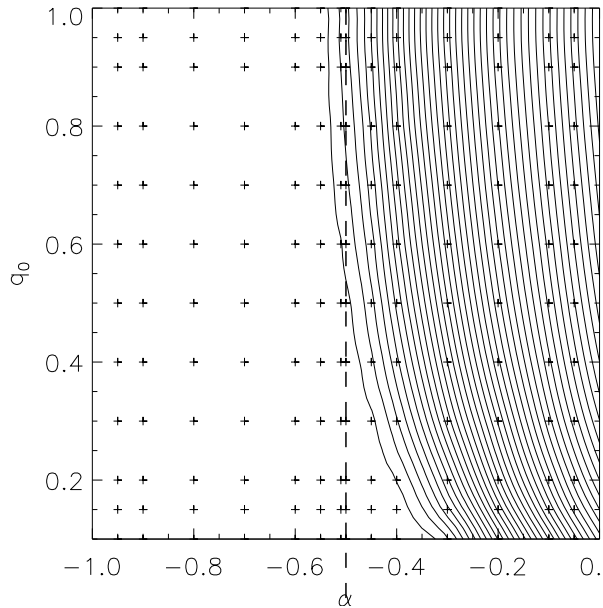,clip=,width=8.cm}} 
\caption{\label{fig:alpha_q} The quality of fit $\chi^2$ as a function
of central cusp slope $\alpha$ and intrinsic flattening $q_0$, for a
value of the flattening at infinity $q_{\infty} = 0.99$. The dashed
line shows the theoretical limit on the cusp slope for two-integral
distribution functions and the crosses indicate combinations of
$\alpha$ and $q_0$ for which models were calculated.}
\end{figure}

We also investigated the axisymmetric version of the triaxial models
with central cusps that were introduced by de Zeeuw \& Carollo
(1996). The potential and density are given in Appendix A. The
behaviour of these models is completely determined by their axis
ratios at small and large radii, $q_0$ and $q_{\infty}$, and the
central cusp slope $\alpha$. The density distribution is dimpled on
the symmetry axis, and, for some choices of $\alpha,q_0$, even becomes
negative. In advance, we can therefore rule out a (small) range of
parameters for which the DF does not exist. The physical range of
$q_0$ decreases when $q_\infty$ is decreased. At $q_{\infty}=0.99$,
which is considered here, models with $q_0 < 0.05$ show this
behaviour. We study the range $0.1 < q_0 < 1$.

We added a central black hole with a mass $\sim 1$\% of the total
mass, and calculated self-consistent two-integral models using the
same grid settings as in \S\ref{sec:kk}. Fig.~\ref{fig:alpha_q} shows
the contours of constant $\chi^2$ as a function of $\alpha$ and
$q_0$. As expected, the $\chi^2$ is zero for values $\alpha<-0.5$ and
rises rapidly above $\alpha=-0.5$, indicating that self-consistent
two-integral distribution functions do not exist at these values of
the central cusp slope. However, for $q_0 < 0.4$, the transition to
$\chi^2 > 0$ occurs at values of $\alpha > -0.5$, which means that
these flatter models have an even larger freedom in possible central
cusp slopes. The points that contribute most to this nonzero $\chi^2$
are located in the center of the model, i.e., near the black hole, and
experiments with a denser sampling in the integrals of motion give
similar results. The limiting behavior of the density is (cf.\
eq~(\ref{rhoPL})
\begin{eqnarray}
\lim_{r \rightarrow 0}\rho(r,\theta) =
     \textstyle{\frac32} g(r)\left[\frac{1}{q^\alpha_0-1}+ \sin^2\theta\right],
\end{eqnarray}
so that the importance of the constant term in the
expansion~(\ref{densjos}) decreases when $q_0$ decreases. This may
gradually relax the restriction on $\alpha$ in the numerical model.  A
similar effect has been found recently in oblate Sridhar-Touma models
(Jalali \& de Zeeuw 2001), which resemble the power-law models in many
respects. However, the values of $q_0$ for which this occurs are much
smaller than the observed flattenings of early-type galaxies.

\section{Summary and conclusions}
\label{sec:summary}

We have presented an alternative to the existing analytical and
numerical methods for calculating two-integral distribution
functions. It is based upon Schwarzschild's orbit superposition method
and able to deal with arbitrary density profiles and galaxy
potentials. Instead of orbital building blocks, we use the so-called
two-integral components, which are smoother and implicitly include
stochastic orbits. Due to their delta-function behaviour, the
observables can be calculated (semi-)analytically, which speeds up the
calculations considerably. We checked that the method is able to
reproduce a known combination of potential, density and distribution
function and tested the regularisation algorithm in the process.  This
test shows that fitting the meridional plane masses alone is not
enough to constrain the DF, while including a (modest) amount of
regularisation is enough to smoothen the DF toward the theoretical
curve.

We have also tested the $\chi^2$ (\ref{eq:chisq}), which measures the
quality-of-fit of the resulting model, against analytical
investigations. This demonstrates that a high value of $\chi^2$
can be taken as a sure indication that the model does not exist.

These results show that our method is flexible and reproduces analytical
results very closely. Additionally, many of the characteristics of the
implementation that we used, such as the regularisation method and the 
$\chi^2$-parameter, are used in more general (three-integral) models. 
Regularisation is therefore also a very useful and necessary tool to 
smoothen the distribution function that is found by these methods, and the 
$\chi^2$-parameter is a useful diagnostic to test the existence of 
the resulting models. Applications in which we calculate fully three-integral
models (including TICs) of observed kinematics will follow in a subsequent paper. 

\medskip

\noindent{\bf Acknowledgements}\\
\noindent It is a pleasure to thank Glenn van de Ven, Michele Cappellari and
Roeland van der Marel for useful discussions and a critical reading of
the manuscript. Furthermore, we are very thankful to Roeland van der Marel
for kindly providing his three-integral Schwarzschild software.

\appendix
\section{A family of axisymmetric models}

The axisymmetric limit of the triaxial models introduced by de Zeeuw
\& Carollo (1996) is defined by the potential
\begin{equation}
V(r,\theta)= -u(r)+ {\textstyle \frac12}(3\cos^2\theta-1)\,v(r),
\end{equation}
with
\begin{eqnarray}
u(r)&=&\left\{\begin{array}{c c}
{\displaystyle \ln \frac{r}{r+1}}&{\rm for}\,\,\alpha=-2\nonumber\\
{\displaystyle \frac{1}{2+\alpha}\,
                   \left[\left(\frac{r}{r+1}\right)^{2+\alpha}-1\right]}&
{\rm for}\,\,\alpha\neq -2\end{array}\right.\nonumber\\
v(r)&=&-\frac{A\,r^{2+\alpha}}{(r+r_2)^{4+\alpha}},
\end{eqnarray}
where $A$ and $R_2$ are constants, we have chosen $GM=1$, and
consider $\alpha \geq -3$. The corresponding density distribution is
\begin{eqnarray}
\label{rhoPL}
\rho(r,\theta)=g(r)\left[\frac{f(r)}{g(r)}-1+{\textstyle \frac32}\sin^2\theta\right]
\end{eqnarray}
with
\begin{eqnarray}
f(r)\!\!\!&=\!\!\!&\frac{(3+\alpha)r^\alpha}
                 {4\pi(r+1)^{4+\alpha}}\nonumber\\
g(r)&=&\frac{A r^\alpha 
              \left[4r^2+4(6+\alpha)r\,r_2-\alpha (5+\alpha)r^2_2\right]}
                                       {4\pi(r+r_2)^{6+\alpha}}.
\end{eqnarray}
The constants $A$ and $r_2$ are related to the axis ratios of the
density distribution at small and large radii, $q_0$ and $q_\infty$,
respectively, by
\begin{eqnarray}
A&=&\frac{(3+\alpha)\,(1-q^4_{\infty})}{2\,(2+q^4_{\infty})}\nonumber\\
r_2&=&\left[\frac{-(1-q^4_{\infty})\,(2+q^{-\alpha}_0)\,\alpha\,(5+\alpha)}
{4\,(1-q^{-\alpha}_0)\,(2+q^4_{\infty})}\right]^{1/{(4+\alpha)}},
\end{eqnarray}

\bsp
\label{lastpage}
\end{document}